\documentstyle[12pt]{article}
\begin{document}
\hfill{NCKU-HEP-98-08}\par
\hfill{hep-ph/9807437}
\vskip 0.5cm
\begin{center}
{\large {\bf New unified evolution equation}}
\vskip 1.0cm
Jyh-Liong Lim
\vskip 0.5cm
Department of Electrophysics, National Chiao-Tung University, \par
Hsinchu, Taiwan 300, Republic of China
\vskip 0.5cm
Hsiang-nan Li
\vskip 0.5cm
Department of Physics, National Cheng-Kung University, \par
Tainan, Taiwan 701, Republic of China
\end{center}
\vskip 1.0cm


%\baselineskip=2\baselineskip
\vskip 1.0cm

\centerline{\bf Abstract}
\vskip 0.3cm

We propose a new unified evolution equation for parton distribution
functions appropriate for both large and small Bjorken variables $x$, which
is an improved version of the Ciafaloni-Catani-Fiorani-Marchesini equation.
In this new equation the cancellation of soft divergences between virtual
and real gluon emissions is explicit without introducing infrared cutoffs
and next-to-leading contributions to the Sudakov resummation can be included
systematically. It is shown that the new equation reduces to the
Dokshitzer-Gribov-Lipatov-Altarelli-Parisi equation at large $x$, to the
Balitsky-Fadin-Kuraev-Lipatov (BFKL) equation at small $x$, and to the
modified BFKL equations with a dependence on momentum transfer $Q$ and with
unitarity, if hard virtual gluon contributions and correction
to strong rapidity ordering are retained, respectively.

\newpage
\centerline{\large\bf 1. Introduction}
\vskip 0.5cm

It is known that the Dokshitzer-Gribov-Lipatov-Altarelli-Parisi (DGLAP)
equation \cite{AP} sums large logarithmic corrections $\ln Q$ to a parton
distribution function at large $x$, with $Q$ being momentum transfer
and $x$ the Bjorken variable, and that the Balitsky-Fadin-Kuraev-Lipatov
(BFKL) equation \cite{BFKL} sums $\ln(1/x)$ at small $x$. The
Ciafaloni-Catani-Fiorani-Marchesini (CCFM) equation \cite{CCFM}, which is
appropriate for both large and small $x$, has been proposed to unify
the above two equations. The conventional derivation of the evolution
equations involves all-order summation of ladder diagrams with rung gluons
obeying specific kinematic orderings. For the DGLAP, BFKL, and CCFM
equations, they are transverse momentum ordering, rapidity ordering, and
angular ordering, respectively.

Recently, we have applied the Collins-Soper (CS) resummation technique
\cite{CS} to all-order summation of various large logarithms \cite{L3}. This
technique was developed originally to organize double logarithms $\ln^2 Q$.
We have demonstrated that it is also applicable to single-logarithm cases,
{\it i.e.}, the derivation of the evolution equations mentioned above
\cite{L3}. In this approach all the evolution kernels are obtained from the
same one-loop diagrams, which are evaluated under appropriate soft
approximations that correspond to specific kinematic orderings. Therefore,
our method is simpler, and provides a unified viewpoint to the evolution
equations. It is easy to modify the approximation employed in the
evaluation of the BFKL kernel, such that a $Q$ dependence can be introduced
into the BFKL equation \cite{L4}, which is important for the explanation of
experimental data, and that predictions from the modified BFKL
equation satisfy the unitarity bound \cite{L5}.

As reproducing the CCFM equation in the framework of resummation,
we notice that real and virtual gluon emissions must be treated in a
different way: the former are computed to lowest order, generating 
a splitting function, while the latter are resummed to all orders,
giving a Sudakov form factor. This observation is consistent with that
in the conventional derivation \cite{CCFM}. However, infrared divergences
must cancel between real and virtual gluon emissions order by order. Hence,
in the conventional derivation infrared cutoffs are necessary for
regularizing the individual divergences of real and virtual 
corrections. These cutoffs may cause an ambiguity, when one intends to
include next-to-leading logarithmic summation into
the evolution equation.

In Ref.~\cite{LL3} we have proposed a new unified evolution equation, which
results from one of the applications of the CS formalism to logarithmic
summations. In this paper we shall present a detailed derivation of
the new equation, explain the ingredients that are added compared to
the CCFM equation, and discuss predictions for the gluon distribution
function. As stated above, infrared divergences of real and virtual gluon
emissions should cancel exactly without introducing artificial cutoffs, and
next-to-leading contributions to the Sudakov resummation can be included
systematically. To achieve these goals, real gluon correction is
separated into two pieces. The piece containing an infrared pole is combined
with the corresponding virtual correction, both of which are resummed into a
Sudakov form factor. The other piece, being infrared finite, is evaluated to
lowest order, giving a splitting function. With such an arrangement, both
the Sudakov form factor and the splitting function are rendered infrared
finite. Using the CS formalism, the running of coupling
constants appearing in the Sudakov form factor and in the splitting function
can be taken into account.

It will be shown that hard virtual gluon contributions introduce a 
dependence on the kinematic variable $Q$ to the rise of the gluon 
distribution function at small $x$. This $Q$ dependence is essential for 
the explanation of experimental data as shown in \cite{L4}. In this respect, 
the new equation, with the $Q$ dependence arising from the $\ln Q$ 
summation, is similar to the CCFM equation. 
It has been observed in \cite{L5} that the assumption of strong rapidity 
ordering overestimates real gluon contributions. This is the reason 
predictions from the BFKL equation violates unitarity constraints. In the 
derivation of the new equation we shall not assume rapidity ordering, and 
obtain a destructive correction to the BFKL rise. With this correction, the
increase of the gluon distribution function becomes mild, and
satisfies the unitarity bound. The underlying mechanism of the above
features are basically the same as that of the modified BFKL equations with
a $Q$ dependence and with unitarity \cite{L4,L5}. We shall argue that
predictions from the CCFM equation violate unitarity constraints.

The CCFM equation is reproduced using the CS resummation technique in
Sec.~2. The new unified evolution equation is derived in Sec.~3. We
demonstrate in Sec.~4 that the new equation reduces to the DGLAP equation,
to the BFKL equation, and to the modified BFKL equations with a $Q$
dependence and with unitarity under different approximations. Section 5
is the conclusion.

\vskip 1.0cm

\centerline{\large\bf 2. CCFM Equation}
\vskip 0.5cm

We study the unintegrated gluon distribution function defined
in the axial gauge $n\cdot A=0$, $n$ being a gauge vector:
\begin{eqnarray}
F(x,k_T,p^+)&=&\frac{1}{p^+}\int\frac{dy^-}{2\pi}\int\frac{d^2y_T}{4\pi}
e^{-i(xp^+y^--{\bf k}_T\cdot {\bf y}_T)}
\nonumber \\
& &\times\frac{1}{2}\sum_\sigma
\langle p,\sigma| F^+_\mu(y^-,y_T)F^{\mu+}(0)|p,\sigma\rangle\;,
\label{deg}
\end{eqnarray}
where $|p,\sigma\rangle$ denotes the incoming hadron with light-like
momentum $p^\mu=p^+\delta^{\mu +}$ and spin $\sigma$, and $F^+_\mu$ is the
field tensor. An average over color is understood. $F(x,k_T,p^+)$ describes
the probability of a gluon carrying a longitudinal momentum fraction $x$
and transverse momenta $k_T$.

For $n^\mu=\delta^{\mu-}$ lying on the light cone, Eq.~(\ref{deg}) is gauge
invariant. This can be understood by considering a variation of $n$ on the
light cone, under which Eq.~(\ref{deg}) is still defined for $A^+=0$ and
remains the same. To derive an evolution equation of a parton
distribution function using the CS formalism, we
must allow $n$ to vary arbitrarily away from the light cone ($n^2\not= 0$),
and a parton distribution function becomes gauge dependent.
However, it will be found that an evolution kernel turns out to be
independent of $n$. This is natural, since it has been proved that
parton distribution functions defined for different $n$ possess the same
infrared structure, and thus the same evolution behavior, though different
ultraviolet structure \cite{L6}. After the derivation, we bring $n$ back to
the light cone, and the gauge invariance of a parton distribution function
is restored \cite{L6}. That is, the vector $n$ appears only at the
intermediate stage of the derivation, and acts as an auxiliary tool.

The key step of the CS formalism is to
obtain the derivative $p^+dF/dp^+$. Because of the scale invariance of $F$ 
in the vector $n$ as indicated by the gluon propagator, 
$-iN^{\mu\nu}(l)/l^2$, with
\begin{equation}
N^{\mu\nu}=g^{\mu\nu}-\frac{n^\mu l^\nu+n^\nu l^\mu}
{n\cdot l}+n^2\frac{l^\mu l^\nu}{(n\cdot l)^2}\;,
\label{gpp}
\end{equation}
$F$ must depend on $p^+$ via the ratio $(p\cdot n)^2/n^2$. Hence,
we have the chain rule relating $p^+d/dp^+$ to $d/dn$ \cite{CS,L1}: 
\begin{eqnarray}
p^+\frac{d}{dp^+}F=-\frac{n^2}{v\cdot n}v_\alpha\frac{d}{dn_\alpha}F\;,
\label{cph}
\end{eqnarray}
with $v^\alpha=\delta^{\alpha +}$ a dimensionless vector along $p$. The
operator $d/dn_\alpha$ applies to a gluon propagator, giving
\cite{CS}
\begin{equation}
\frac{d}{dn_\alpha}N^{\mu\nu}=
-\frac{1}{n\cdot l}(l^\mu N^{\alpha\nu}+l^\nu N^{\mu\alpha})\;.
\label{dgp}
\end{equation}
The loop momentum $l^\mu$ ($l^\nu$) in the above expression contracts with
a vertex the differentiated gluon attaches, which is then replaced by a
special vertex \cite{CS,L1}
\begin{eqnarray}
{\hat v}_\alpha=\frac{n^2v_\alpha}{v\cdot nn\cdot l}\;.
\label{va}
\end{eqnarray}
This special vertex can be easily read off from the combination of 
Eqs.~(\ref{cph}) and (\ref{dgp}).

Employing the Ward identities for the contraction of $l^\mu$ ($l^\nu$)
\cite{L3}, the sum of diagrams with different differentiated gluons
reduces to a new diagram, in which the special vertex moves to the outer
end of a parton line \cite{CS,L1}. We obtain the formula
\begin{equation}
p^+\frac{d}{dp^+}F(x,k_T,p^+)=2{\bar F}(x,k_T,p^+)\;,
\label{cc}
\end{equation}
shown in Fig.~1(a), where $\bar F$ denotes the new diagram, and the 
square represents the special vertex. The coefficient 2 comes from the
equality of the new functions with the special vertex on either of the two
parton lines.

The leading regions of the loop momentum flowing through the special
vertex are soft and hard, since the vector $n$ does not lie on the light
cone, and collinear enhancements are suppressed. If following the
standard resummation procedure, we should factorize out subdiagrams
containing the special vertex order by order in the leading regions 
(see Figs.~1(b) and 1(c)). This procedure will lead to a new unified 
evolution equation in the next section. To reproduce the CCFM 
equation, however, the factorization must be performed in a different way
described by Fig.~2(a), where the two jet functions $J$ group all-order 
virtual corrections, and the lowest-order real gluon emission between them 
is soft. This factorization picture is motivated by the conventional 
derivation in \cite{CCFM}.

We first resum the double logarithms contained in $J$ by considering its
derivative
\begin{eqnarray}
p^+\frac{d}{dp^+}J(p_T,p^+)&=&{\bar J}(k_T,p^+)
\nonumber \\
&=&[K_J(k_T/\mu,\alpha_s(\mu))+G_J(p^+/\mu,\alpha_s(\mu))]J(k_T,p^+)\;,
\label{ccj}
\end{eqnarray}
which is similar to Eq.~(\ref{cc}). In the second expression subdiagrams
containing the special vertex have been factorized into the functions $K_J$
and $G_J$ in the leading soft and hard regions, respectively. At lowest
order, $K_J$ comes from Fig.~2(b), where the replacement of the
gluon line by an eikonal line will be explained below, and $G_J$ comes from
Fig.~2(c). The second diagram in Fig.~2(c) serves as a soft subtraction,
which guarantees a hard momentum flow in $G_J$. We have set the infrared
cutoff of $K_J$ to $k_T$, as indicated by its argument, in order to
regularize its soft pole. This cutoff is necessary here due to the lack of
the corresponding real gluon emission diagram.
At this step, an ambiguity in the
summation of single logarithms is introduced.

We evaluate the contribution from Fig.~2(b). The color factor is
extracted from the relation $f_{abc}f_{bdc}=-N_c\delta_{ad}$, where the
indices $a,b,\dots$ are referred to Fig.~1(b), and $N_c=3$ is the
number of colors. The one-loop $K_J$ is written as
\begin{eqnarray}
K_J=iN_cg^2\int\frac{d^{4}l}{(2\pi)^4}N_{\nu\beta}(l)
\frac{{\hat v}^\beta \Gamma^{\mu\nu\lambda}}{-2xp\cdot l l^2}
-\delta K_J\;,
\label{kj1}
\end{eqnarray}
with $\delta K_J$ an additive counterterm. The triple-gluon vertex is
\begin{eqnarray}
\Gamma^{\mu\nu\lambda}&=&g^{\mu\nu}
(xp+l)^{\lambda}+g^{\nu\lambda}(xp-2l)^{\mu}+g^{\lambda\mu}
(l-2xp)^{\nu}\;,
\label{tri}\\
&\approx& xp^+(g^{\mu\nu}v^{\lambda}+
g^{\nu\lambda}v^{\mu}-2g^{\lambda\mu}v^{\nu})\;,
\label{tris}
\end{eqnarray}
where Eq.~(\ref{tris}) is the soft loop momentum ($l\to 0$) limit
of Eq.~(\ref{tri}). The factor $1/(-2xp\cdot l)$ is the soft
approximation of the gluon propagator $1/(l-xp)^2$.

It has been shown that the terms $v^\lambda$ and $v^\mu$ in
Eq.~(\ref{tris}) result in contributions suppressed by a power $1/s$,
$s=(p+q)^2$, compared to the contribution from the last term $v^\nu$
\cite{L3,L4}. Absorbing the metric tensor $g^{\lambda\mu}$
into $F$, Eq.~(\ref{kj1}) is simplified into
\begin{eqnarray}
K_J=iN_cg^2\int\frac{d^{4}l}{(2\pi)^4}N_{\nu\beta}(l)
\frac{{\hat v}^\beta v^{\nu}}{v\cdot l l^2}-\delta K_J\;.
\label{kj2}
\end{eqnarray}
The factor $v^\nu/(v\cdot l)$ is represented by an eikonal line in
Fig.~2(b), with the Feynman rules $v^\nu$ for an eikonal vertex and
$1/(v\cdot l)$ for an eikonal propagator.

Since ultraviolet poles exist in Fig.~2(b) and in the second diagram
of Fig.~2(c), the sum $K_J+G_J$ is independent of $\mu$, {\it i.e.},
renormalization-group (RG) invariant. The standard RG analysis leads to
\begin{equation}
\mu\frac{d}{d\mu}K_J=-\gamma_J=-\mu\frac{d}{d\mu}G_J\;,
\end{equation}
whose solution is written as
\begin{equation}
K_J(k_T/\mu,\alpha_s(\mu))+G_J(p^+/\mu,\alpha_s(\mu))=
-\gamma_J(\alpha_s)\int_{k_T}^{p^+}\frac{d{\bar\mu}}{\bar\mu}\;.
\label{cckg}
\end{equation}
The anomalous dimension $\gamma_J$ is found to be \cite{L3}
\begin{equation}
\gamma_J=\mu \frac{d}{d\mu}\delta K_J={\bar\alpha_s}\;,
\end{equation}
with the coupling constant ${\bar\alpha_s}=N_c\alpha_s/\pi$.
To the accuracy of leading (double) logarithms \cite{CCFM}, we have
neglected the initial conditions $K_J(1,\alpha_s(k_T))$ and 
$G_J(1,\alpha_s(p^+))$ of the RG evolution, and the running of
${\bar\alpha}_s$, which contribute to
next-to-leading (single) logarithmic summation. In the derivation of the
new unified evolution equation these factors can be included.

Substituting Eq.~(\ref{cckg}) into Eq.~(\ref{ccj}), we solve for 
\begin{equation}
J(k_T,Q)=\Delta^{1/2}(k_T,Q)J^{(0)}\;,
\label{jd}
\end{equation}
with the double-logarithm exponential 
\begin{eqnarray}
\Delta(k_T,Q)=\exp\left[-2{\bar\alpha_s}
\int_{k_T}^{Q}\frac{dp^+}{p^+}
\int_{k_T}^{p^+}\frac{d{\bar\mu}}{\bar\mu}\right]\;.
\label{db}
\end{eqnarray}
where the upper bound of $p^+$ has been chosen as $Q$. The initial 
condition $J^{(0)}$ can be regarded as a
tree-level gluon propagator, which will be eikonalized in the evaluation of
the lowest-order soft real gluon emission below.

We split the above exponential into
\begin{equation}
\Delta(k_T,Q)=\Delta_S(zq,Q)\Delta_{NS}(z,q,k_T)\;, 
\label{split}
\end{equation}
with $z=x/\xi$ and $q=l_T/(1-z)$, where $\xi$ is the momentum fraction
entering $J$ from the bottom, and $l_T$ is the transverse loop momentum 
carried by the soft real gluon. The so-called ``Sudakov" exponential
$\Delta_S$ and the ``non-Sudakov" exponential $\Delta_{NS}$ are given by
\begin{eqnarray}
\Delta_S(zq,Q)&=&\exp\left[-2{\bar\alpha_s}
\int_{zq}^{Q}\frac{dp^+}{p^+}
\int_{k_T}^{p^+}\frac{d{\bar\mu}}{\bar\mu}\right]\;,
\nonumber \\
&=&\exp\left[-{\bar\alpha_s}
\int_{(zq)^2}^{Q^2}\frac{dp^2}{p^2}
\int_{0}^{1-k_T/p}\frac{dz'}{1-z'}\right]\;,
\label{dbs} \\
\Delta_{NS}(z,q,k_T)&=&\exp\left[-2{\bar\alpha_s}
\int_{k_T}^{zq}\frac{dp^+}{p^+}
\int_{k_T}^{p^+}\frac{d{\bar\mu}}{\bar\mu}\right]\;,
\nonumber \\
&=&\exp\left[-{\bar\alpha_s}\int_{z}^{k_T/q}\frac{dz'}{z'}
\int_{(z'q)^2}^{k_T^2}\frac{dp^2}{p^2}\right]\;,
\label{nons}
\end{eqnarray}
where the variable changes ${\bar \mu}=(1-z')p$ and $p^+=p$ for $\Delta_S$, 
and ${\bar \mu}=p$ and $p^+=z'q$ for $\Delta_{NS}$ have been adopted to
obtain the second expressions. 

Using Eq.~(\ref{jd}) for the two jet functions, $\bar F$ from Fig.~2(a) is 
written as
\begin{eqnarray}
{\bar F}(x,k_T,p^+)&=&iN_cg^2\int\frac{d^4l}{(2\pi)^4}
N_{\nu\beta}(l)\frac{{\hat v}^\beta v^\nu}{v\cdot l}
2\pi i\delta(l^2)\Delta(k_T,Q)
\nonumber \\
& &\times \theta(Q-zq)
F(x+l^+/p^+,|{\bf k}_T+{\bf l}_T|,p^+)\;,
\label{ci1}
\end{eqnarray}
where the tree-level gluon propagator $J^{(0)}$ on the right-hand side has
been eikonalized to give the factor $1/v\cdot l$, and that on the left-hand 
side has been absorbed into $F$. The extra $\theta$ function,
requiring $Q>zq$, renders the Sudakov exponential $\Delta_S$ meaningful. This 
relation comes from the angular ordering of radiative gluons,
$Q/(xp^+) > l_T/[(\xi-x)p^+]$. Compared with the transverse momentum 
ordering for the DGLAP equation, $Q$ ($l_T$) has been divided by the 
longitudinal momentum $xp^+$ ($(\xi-x)p^+$). Therefore, the inserted scale 
$zq$ reflects the specific kinematic ordering for the CCFM equation. Those 
radiative gluons, which do not obey angular ordering, contribute to the 
non-Sudakov exponential.

Adopting the variable change $\xi=x+l^+/p^+$ and performing the integration
over $l^-$, we obtain
\begin{eqnarray}
{\bar F}(x,k_T,p^+)&=&\frac{\bar\alpha_s}{2}\int_x^1 d\xi
\int\frac{d^2l_T}{\pi}
\frac{2n^2(\xi-x)p^{+2}}{[n^+l_T^2+2n^-(\xi-x)^2p^{+2}]^2}\Delta(k_T,Q)
\nonumber\\ 
& &\times \theta(Q-zq)F(\xi,|{\bf k}_T+{\bf l}_T|,p^+)\;,
\label{cctf}
\end{eqnarray}
where $n=(n^+,n^-,{\bf 0})$ has been chosen for convenience. The above
expression is then substituted into Eq.~(\ref{cc}) to find a solution of
$F$. Integrating Eq.~(\ref{cctf}) from $p^+=0$ to $Q$, and applying the
variable changes $\xi=x/z$ and ${\bf l}_T=(1-z){\bf q}$, we have
\begin{eqnarray}
F(x,k_T,Q)&=&F^{(0)}+{\bar\alpha_s}\int_x^1 dz
\int\frac{d^2q}{\pi q^2}\theta(Q-zq)
\Delta_S(Q,zq)\Delta_{NS}(z,q,k_T)
\nonumber \\
& &\hspace{2.0cm}\times
\frac{1}{z(1-z)}F(x/z,|{\bf k}_T+(1-z){\bf q}|,q)\;,
\label{ccc1}
\end{eqnarray}
where the initial condition $F^{(0)}$ corresponds to the 
lower bound of $p^+$. To work out the $p^+$ integration,
$F(x/z,|{\bf k}_T+{\bf l}_T|,p^+)$ in Eq.~(\ref{cctf}) has been
approximated by $F(x/z,|{\bf p}_T+{\bf l}_T|,q)$. Note that the $n$
dependence has disappeared, and the CCFM kernel is gauge invariant. 

Equation (\ref{ccc1}) can be reexpressed as
\begin{eqnarray}
F(x,k_T,Q)&=&F^{(0)}+{\bar\alpha}_s\int_x^1 dz
\int\frac{d^2q}{\pi q^2}\theta(Q-zq)
\Delta_S(zq,Q){\tilde P}(z,q,k_T)
\nonumber \\ 
& &\hspace{2.0cm}\times
F(x/z,|{\bf k}_T+(1-z){\bf q}|,q)\;,
\label{ccfm}
\end{eqnarray}
with the function
\begin{equation}
{\tilde P}=
\left[\frac{1}{(1-z)_+}+\Delta_{NS}(z,q,k_T)\frac{1}{z}-2+z(1-z)\right]\;,
\label{tp}
\end{equation}
which is close to the splitting function
\begin{equation}
P_{gg}=\left[\frac{1}{(1-z)_+}+\frac{1}{z}-2+z(1-z)\right]\;.
\label{pgg}
\end{equation}
Equation (\ref{ccfm}) is the CCFM equation \cite{CCFM}. To arrive at
Eq.~(\ref{tp}), we have employed the identity
$1/[z(1-z)]\equiv 1/(1-z) +1/z$, and put in by hand the last term
$-2+z(1-z)$. This term, finite at $z\to 0$ and at $z\to 1$, can not be
obtained in \cite{CCFM} either due to a soft approximation. Note that
Eq.~(\ref{ccc1}) contains an infrared singularity from $z\to 1$. Hence,
we have replaced the factor $1/(1-z)$ by the plus distribution $1/(1-z)_+$
in Eq.~(\ref{ccfm}), and dropped the soft pole corresponding
to $z\to 1$. On the other hand, only the non-Sudakov form factor
$\Delta_{NS}$ in front of $1/z$ is retained. Because $\Delta_{NS}$ vanishes
when the upper bound $zq$ of $p^+$ approaches zero, as indicated by
Eq.~(\ref{nons}), it smears the $z\to 0$ pole in ${\tilde P}$ \cite{CCFM}.

As shown above, the double-logarithm exponential $\Delta$ and the function
$\tilde P$ are infrared divergent individually. To have a well-defined
evolution equation, the infrared regularizations for $\Delta$ and $\tilde P$
must be implemented. The lower bound of the variable $z$ is $x$, and never
goes down to zero. Therefore, the smearing factor $\Delta_{NS}$ and the
splitting of the double-logarithm exponential in Eq.~(\ref{split}) are not
essential. The new unified evolution equation derived in the next section
will take into account these observations.

\vskip 1.0cm

\centerline{\large\bf 3. New Unified Equation}
\vskip 0.5cm

In this section we shall derive a new unified evolution equation following
the standard procedure of the CS formalism. We start with Eq.~(\ref{cc})
for the unintegrated gluon distribution function
$F(x,k_T,p^+)$. We Fourier transform Eq.~(\ref{cc}) into $b$
space with $b$ being the conjugate variable of $k_T$. In the leading soft
and hard regions of the loop momentum flowing through the special vertex,
$\bar F$ can be factorized into the convolution of subdiagrams
containing the special vertex with the original distribution function $F$.
To lowest order, the soft contribution from the subdiagrams in Fig.~1(b) is
written as ${\bar F}_s={\bar F}_{sv}+{\bar F}_{sr}$ with
\begin{eqnarray}
{\bar F}_{sv}&=&iN_cg^2\int\frac{d^4l}{(2\pi)^4}
N_{\nu\beta}(l)\frac{{\hat v}^\beta v^\nu}{v\cdot l l^2}F(x,b,p^+)\;,
\label{fsv} \\
{\bar F}_{sr}&=&iN_cg^2\int\frac{d^4l}{(2\pi)^4}
N_{\nu\beta}(l)\frac{{\hat v}^\beta v^\nu}{v\cdot l}
2\pi i\delta(l^2)e^{i{\bf l}_T\cdot {\bf b}}
\nonumber\\
& &\times F(x+l^+/p^+,b,p^+)\;,
\label{fsr}
\end{eqnarray}
corresponding to virtual and real gluon emissions, respectively. 
Equation (\ref{fsv}) is the same as the integral in Eq.~(\ref{kj2}).
The factor $\exp(i{\bf l}_T\cdot {\bf b})$ in Eq.~(\ref{fsr}) comes from 
Fourier transformation of the real gluon contribution.

We reexpress the function $F$ in the integrand of ${\bar F}_{sr}$ as
\begin{eqnarray}
& &F(x+l^+/p^+,b,p^+)=\theta((1-x)p^+-l^+)F(x,b,p^+)
\nonumber\\
& &\hspace{1.5cm} +[F(x+l^+/p^+,b,p^+)-\theta((1-x)p^+-l^+)F(x,b,p^+)]\;.
\label{fre}
\end{eqnarray}
The contribution from the first term on the right-hand 
side of Eq.~(\ref{fre}) is combined with ${\bar F}_{sv}$, leading to
$K(b\mu,\alpha_s(\mu))F(x,b,p^+)$ with the function
\begin{eqnarray}
K&=&iN_cg^2\int\frac{d^4l}{(2\pi)^4}
N_{\nu\beta}(l)\frac{{\hat v}^\beta v^\nu}{v\cdot l}
\nonumber \\
& & \times \left[\frac{1}{l^2}
+2\pi i\delta(l^2)\theta((1-x)p^+-l^+)e^{i{\bf l}_T\cdot {\bf b}}\right]
-\delta K\;,
\label{fs1}
\end{eqnarray}
where $\delta K$ is an additive counterterm.  The 
contribution from the second term in Eq.~(\ref{fre}) is denoted as 
${\bar F}'_{s}(x,b,p^+)$,
\begin{eqnarray}
{\bar F}'_{s}&=&iN_cg^2\int\frac{d^4l}{(2\pi)^4}
N_{\nu\beta}(l)\frac{{\hat v}^\beta v^\nu}{v\cdot l}
2\pi i\delta(l^2)e^{i{\bf l}_T\cdot {\bf b}}
\nonumber \\
& & \times[F(x+l^+/p^+,b,p^+)-\theta((1-x)p^+-l^+)F(x,b,p^+)]\;.
\label{fs2}
\end{eqnarray}
That is, we have separated the real gluon contribution into two pieces. The 
piece with an infrared pole, along with the virtual gluon contribution, goes 
into the function $K$. Compared with $K_J$ in Eq.~(\ref{kj2}), $K$ is
infrared finite due to the inclusion of part of the real correction. The
other piece ${\bar F}'_s$, being also infrared finite, will lead to a
splitting function below.

Fig.~1(c) gives $G(k^+/\mu,\alpha_s(\mu))F(x,b,p^+)$ with the function
\begin{eqnarray}
G=iN_cg^2\int\frac{d^4l}{(2\pi)^4}
N_{\nu\beta}(l)\frac{{\hat v}^\beta}{l^2}
\left[\frac{(l-2k^+v)^{\nu}}{(l-k^+v)^2}-\frac{v^\nu}{v\cdot l}\right]
-\delta G\;,
\label{gpb}
\end{eqnarray}
where $k^+=xp^+$ is the parton momentum and
$\delta G$ an additive counterterm.
To obtain the above expression, Eq.~(\ref{tri}) for the triple-gluon
vertex has been inserted, which generates the first term in the brackets.
At intermediate $x$, $G$ is characterized by a large scale $k^+$, 
whose logarithms $\ln k^+$ imply that collinear divergences are present
in $F$, and that the Sudakov resummation of double logarithms from the
overlap of collinear and soft enhancements is necessary. Through this
function $G$, the dependence on a large scale is introduced into the
new evolution equation as stated before.

The functions $K$ and $G$ contain ultraviolet divergences individually as
indicated by the renormalization scale $\mu$. However, their ultraviolet
divergences, both from the virtual contribution ${\bar F}_{sv}$,
cancel each other. This is obvious from the counterterms
\begin{equation}
\delta K=-\delta G={\bar\alpha}_s\left(\frac{1}{\epsilon}
+\frac{1}{2}\ln 4\pi-\frac{\gamma_E}{2}\right)\;,
\end{equation}
with the Euler constant $\gamma_E$. Therefore, the sum $K+G$ is RG
invariant, the same as $K_J+G_J$ in Eq.~(\ref{cckg}). The results of $K$
and $G$ from Eqs.~(\ref{fs1}) and (\ref{gpb}) are
\begin{eqnarray}
K(b\mu,\alpha_s(\mu))&=&-{\bar \alpha}_s(\mu)\left[
K_0(2(1-x)p^+\nu b)+\ln \frac{b\mu}{2}+\gamma_E\right]\;,
\nonumber \\
G(k^+/\mu,\alpha_s(\mu))&=&-{\bar \alpha}_s(\mu)\left[
\ln\frac{2k^+}{\mu}+\ln\nu\right]\;,
\end{eqnarray}
with the gauge factor $\nu=\sqrt{(v\cdot n)^2/|n^2|}$.

The RG solution of $K+G$ is given by
\begin{eqnarray}
& &K(b\mu,\alpha_s(\mu))+G(k^+/\mu,\alpha_s(\mu))=
\nonumber \\
& &\hspace{1.0 cm}K(1,\alpha_s(k^+))+G(1,\alpha_s(k^+))-s(b,k^+)\;,
\label{shkg}
\end{eqnarray}
with
\begin{eqnarray}
& &K(1,\alpha_s(k^+))={\bar \alpha}_s(k^+)\left[\ln(1-x)+\ln(p^+b)+
\ln 2\nu\right]\;,
\label{ka}\\
& &G(1,\alpha_s(k^+))=-{\bar \alpha}_s(k^+)\ln 2\nu\;,
\label{gin}\\
& &s(b,k^+)=\int_{1/b}^{k^+}\frac{d{\bar\mu}}{\bar\mu}
\left[\gamma_K(\alpha_s({\bar\mu}))
+\beta(g)\frac{\partial}{\partial g}K(1,\alpha_s({\bar\mu}))\right]\;,
\label{se}
\end{eqnarray}
$\gamma_K=\mu d \delta K/d\mu$ being the anomalous dimension of $K$ and
$\beta(g)$ the beta function.
To obtain Eq.~(\ref{ka}), we have kept only the first two terms in the
power series of the Bessel function $K_0(x)\approx -\ln (x/2)-\gamma_E$,
which is a reasonable approximation at small $b$. We shall show that the 
large $b$ region is suppressed by the Sudakov form factor from 
double-logarithm resummation. Even if there is no Sudakov suppression,
the intrinsic $b$ dependence of the gluon distribution function
$F(x,b,p^+)$, usually parametrized as a Gaussian form \cite{KLM}, still 
provides suppression. 

The one-loop expression of $\gamma_K$ in Eq.~(\ref{se}) gives leading
(double) logarithmic summation, and the two-loop expression gives 
next-to-leading (single) logarithmic summation. Another term proportional
to $\beta\partial K/\partial g$ and the initial conditions $K(1,\alpha_s)$
and $G(1,\alpha_s)$ also give next-to-leading contributions. Hence, the
Sudakov resummation can be performed up to the accuracy of single logarithms,
and the running of the coupling constant must be taken into account in our
formalism. Note that it is equivalent to choose $Ck^+$ as the upper bound of
${\bar\mu}$ in Eq.~(\ref{se}), where $C$ a constant of order unity. The
ambiguity from this arbitrary constant $C$ is cancelled by the ambiguity
from the initial condition $G(1/C,\alpha_s)$ at the single-logarithm
level. The terms involving the
gauge factor $\ln 2\nu$ in the initial conditions of $K$ and $G$ cancel. 

For ${\bar F}'_{s}$, we employ the variable change $\xi=x+l^+/p^+$, and
perform the integration straightforwardly, obtaining
\begin{eqnarray}
{\bar F}'_{s}&=&2{\bar\alpha}_s(k^+)p^+\nu b\int_x^1
d\xi K_1\left(2(\xi-x)p^+\nu b\right)
\nonumber\\
& &\times\left[F(\xi,b,p^+)-F(x,b,p^+)\right]\;,
\nonumber\\
&\approx &{\bar\alpha}_s(k^+)\int_x^1 \frac{d\xi}{\xi}
\frac{F(\xi,b,p^+)-F(x,b,p^+)}{1-x/\xi}\;.
\label{sf2}
\end{eqnarray}
For a similar reason, we have taken the small $b$ limit of the Bessel
function $K_1$. At this stage, the gauge dependence from $\nu$ in the
evolution kernel disappears completely. The argument of ${\bar\alpha}_s$ in
the above expression has been chosen as $k^+$ according to Eq.~(\ref{shkg}),
since ${\bar F}'_{s}$ is a part of the function $K$. Further applying the
variable change $\xi=x/z$ and using the identity
$1/[z(1-z)]\equiv 1/(1-z) +1/z$, Eq.~(\ref{sf2}) is rewritten as
\begin{eqnarray}
{\bar F}'_{s}={\bar\alpha}_s(k^+)\int_x^1dz\left
(\frac{1}{z}+\frac{1}{1-z}\right)
\left[F(x/z,b,p^+)-F(x,b,p^+)\right]\;.
\label{sf3}
\end{eqnarray}

Combining Eqs.~(\ref{shkg}) and (\ref{sf3}), the evolution equation becomes
\begin{eqnarray}
p^+\frac{d}{dp^+}F(x,b,p^+)&=&-2\left[s(b,k^+)-{\bar \alpha}_s(k^+)\ln(k^+b)
\right]F(x,b,p^+)
\nonumber\\
& &+2{\bar\alpha}_s(k^+)\int_x^1 dz P_{gg}(z)F(x/z,b,p^+)\;,
\label{ue2}
\end{eqnarray}
with $P_{gg}$ defined by Eq.~(\ref{pgg}).  The term ${\bar \alpha}_s\ln(1-x)$
in Eq.~(\ref{ka}) has been incorporated with $1/(1-z)$ in Eq.~(\ref{sf3}) to
form the plus distribution $1/(1-z)_+$ appearing in a DGLAP splitting
function. The term $-\int dz F(x,b,p^+)/z$ in Eq.~(\ref{sf3}) has been
combined with $\ln(p^+b)$ in Eq.~(\ref{ka}), leading to 
$\ln(k^+b)$ in Eq.~(\ref{ue2}. The terms finite at
$z\to 0$ and at $z\to 1$ in $P_{gg}$ are put in by hand as in derivation
of the CCFM equation \cite{CCFM}.
In the derivation below we shall ignore the term 
${\bar \alpha}_s(k^+)\ln(k^+b)$ in Eq.~(\ref{ue2}), which is 
less important than $s(b,k^+)$. The integral of $s$ gains a large 
contribution from the low end of the integration variable
${\bar\mu}\sim 1/b$, since the running coupling constant
${\bar \alpha}_s(1/b)$ is much greater than ${\bar \alpha}_s(k^+)$
in the large $b$ region.

The Sudakov form factor $\Delta$ can be extracted by assuming
\begin{equation}
F(x,b,p^+)=\Delta(x,b,p^+,Q_0)F^{(s)}(x,b,p^+)\;,
\label{fsf}
\end{equation}
with the exponential
\begin{eqnarray}
\Delta(x,b,p^+,Q_0)=\exp\left[-2\int_{Q_0}^{p^+}\frac{d\mu}{\mu}
s(b,x\mu)\right]\;,
\label{del}
\end{eqnarray}
where $Q_0$ is an arbitrary low energy scale, satisfying
$xQ_0\ge 1/b$. As $xQ_0<1/b$, implying the vanishing of collinear
enhancements, the Sudakov form factor should be set to unity. 
Equation (\ref{del}) indicates that the Sudakov form factor exhibits 
strong suppression at large $b$, and that the small $b$ expansion in the 
derivation of Eqs.~(\ref{ka}) and (\ref{sf2}) makes sense.

Substituting Eq.~(\ref{fsf}) into Eq.~(\ref{ue2}), we have the
differential equation of $F^{(s)}$,
\begin{eqnarray}
p^+\frac{d}{dp^+}F^{(s)}(x,b,p^+)=\frac{2{\bar\alpha}_s(k^+)}
{\Delta(x,b,p^+,Q_0)}
\int_x^1 dz P_{gg}(z)F(x/z,b,p^+)\;.
\label{ue3}
\end{eqnarray}
The above equation is rewritten as
\begin{eqnarray}
F^{(s)}(x,b,Q)=F^{(s)}(x,b,Q_0)
+\int_x^1 dz\int_{Q_0^2}^{Q^2}\frac{d\mu^2}{\mu^2}
\frac{{\bar\alpha}_s(x\mu)P_{gg}(z)}{\Delta(x,b,\mu,Q_0)}F(x/z,b,\mu)\;,
\label{ue4}
\end{eqnarray}
where $F^{(s)}(x,b,Q_0)=F(x,b,Q_0)$ is the initial condition of
$F$ at the low scale $p^+=Q_0$. We have chosen the upper bound of $p^+$ as 
the measured kinematic variable $Q$. Finally, using Eq.~(\ref{fsf}),
the unintegrated gluon distribution function in $b$ space is given by
\begin{eqnarray}
F(x,b,Q)&=&\Delta(x,b,Q,Q_0)F(x,b,Q_0)
\nonumber\\
& &+\int_x^1 dz\int_{Q_0^2}^{Q^2}\frac{d\mu^2}{\mu^2}
{\bar\alpha}_s(x\mu)\Delta(x,b,Q,\mu)P_{gg}(z)F(x/z,b,\mu)\;,
\nonumber\\
& &
\label{nunif}
\end{eqnarray}
Equation (\ref{nunif}) is the new unified evolution equation.

\vskip 1.0cm
\centerline{\large\bf 4. Features of the New Equation}
\vskip 0.5cm

In this section we shall explore features of the new unified evolution 
equation. We first demonstrate that the new equation
approaches the DGLAP equation at large $x$, The strong $k_T$ ordering of
radiative gluons corresponds to small transverse separation $b$ among these
gluons. In the region with small $b$ and large $x$, we have
\begin{eqnarray}
& &{\bar\alpha}_s(x\mu)\approx {\bar\alpha}_s(\mu)\;,
\label{xap}\\
& &\lim_{b\to 1/k^+\to 0} s(b,k^+)=0\;,
\label{lsu}\\
& &\lim_{b\to 0}F(x,b,Q)=xG(x,Q)\;.
\label{ftg}
\end{eqnarray}
Equation (\ref{lsu}) corresponds to the fact that the effect of Sudakov
resummation is weak at small $b$ as explained in Sec.~3.
Equation (\ref{ftg}) comes from the definition of the gluon density $G(x,Q)$
as the integral of $F(x,k_T,Q)$ in Eq.~(\ref{deg}) over $k_T$:
\begin{eqnarray}
\lim_{b\to 0}F(x,b,p^+)&=&\int\frac{d^2k_T}{\pi}F(x,k_T,p^+)
\nonumber\\
&=&\frac{1}{p^+}\int\frac{dy^-}{2\pi}
e^{-ixp^+y^-}\int d^2y_T\delta^2({\bf y}_T)
\nonumber \\
& &\times\frac{1}{2}\sum_\sigma
\langle p,\sigma| F^+_\mu(y^-,y_T)F^{\mu+}(0)|p,\sigma\rangle
\nonumber\\
&=&xG(x,p^+)\;.
\label{dep}
\end{eqnarray} 
Differentiating Eq.~(\ref{nunif}) with respect to $Q^2$ under 
Eqs.~(\ref{xap})-(\ref{ftg}), we obtain
\begin{eqnarray}
Q^2\frac{d}{d Q^2}G(x,Q)={\bar\alpha}_s(Q)\int_x^1 \frac{dz}{z}
P_{gg}(z)G(x/z,Q)\;,
\label{ap}
\end{eqnarray}
which is the DGLAP equation for $G(x,Q)$. The initial condition
$F(x,b,Q_0)$ does not depend on $Q$, and thus its derivative vanishes.

We then show that the new evolution equation reduces to the BFKL equation
in the small-$x$ limit. For convenience, we shall adopt the one-loop 
expression $\gamma_K={\bar\alpha}_s$, which is the same as $\gamma_J$ for 
the CCFM equation, and ignore the running of ${\bar\alpha}_s$ below. Strong 
rapidity ordering corresponds to the assumption that the $z$ integral in 
the second term of Eq.~(\ref{nunif}) is dominated by the behavior of 
$F(x/z,b,\mu)$ at small $x/z$, {\it i.e.}, at $z\gg x$ \cite{L3}.
Substituting $F(x,b,Q)$ for $F(x/z,b,Q)$, Eq.~(\ref{nunif}) becomes
\begin{eqnarray}
F(x,b,Q)&=&\Delta(x,b,Q,Q_0)F(x,b,Q_0)
\nonumber\\
& &+{\bar\alpha}_s\int_x^1 \frac{dz}{z}\int_{Q_0^2}^{Q^2}
\frac{d\mu^2}{\mu^2}\Delta(x,b,Q,\mu)F(x,b,\mu)\;.
\label{nbk}
\end{eqnarray}
In the above expression the term $-2+z(1-z)$ in the splitting function
$P_{gg}$, which is finite as $z\to x\to 0$, has been
neglected. 

In the small $x$ limit we have
\begin{eqnarray}
& &\Delta(x,b,Q,Q_0)\approx 1\;,
\\
& &-x\frac{d}{dx}\Delta(x,b,Q,Q_0)
={\bar\alpha}_s\ln\frac{Q^2}{Q_0^2}\Delta(x,b,Q,Q_0)
\approx {\bar\alpha}_s\ln\frac{Q^2}{Q_0^2}\;.
\label{sde}
\end{eqnarray}
Differentiating $F$ with respect to $x$ under the above relations,
Eq.~(\ref{nbk}) leads to
\begin{eqnarray}
-x\frac{d}{d x}F(x,b,Q)&=&
{\bar\alpha}_s\int_{Q_0^2}^{Q^2}\frac{d\mu^2}{\mu^2}F(x,b,\mu)\;,
\nonumber\\
&\approx & {\bar\alpha}_s\ln\frac{Q^2}{Q_0^2}F(x,b,Q)\;.
\label{mbf}
\end{eqnarray}
The derivative of the first term on the right-hand side of Eq.~(\ref{nbk}) 
has been neglected, since $F^{(0)}$ is assumed
to be ``flat" \cite{KLM}. The derivatives of $F(x,b,\mu)$ and of
$\Delta(x,b,Q,\mu)$ in the integrand
of the second term have been also neglected, because they give 
next-to-leading-order contributions. To derive the second expression, we
have made the approximation $F(x,b,\mu)\approx F(x,b,Q)$, which maintains
the feature of the evolution in $x$. Equation (\ref{mbf}) is the modified 
BFKL equation with a $Q$ dependence derived in \cite{L4}.

If we drop hard virtual gluon contributions, the $Q$ dependence will
not appear in the evolution kernel. Replacing $Q$ 
by $1/b$, which is the only scale, Eq.~(\ref{mbf}) reduces to
\begin{eqnarray}
-x\frac{d}{d x}F(x,b)= -{\bar\alpha}_s\ln(Q_0^2 b^2)F(x,b)\;.
\label{bfkl}
\end{eqnarray}
Since the evolution kernel does not depend on $Q$, 
the argument $Q$ of $F$ has been suppressed. It can be shown that the 
Fourier transformation of the BFKL equation adopted in \cite{L4,KLM},
\begin{eqnarray}
-x\frac{d}{d x}F(x,k_T)={\bar\alpha}_s
\int\frac{d^2l_T}{\pi l_T^2}[F(x,|{\bf k}_T+{\bf l}_T|)-
\theta(Q_0-l_T)F(x,k_T)]\;,
\label{bfk}
\end{eqnarray}
coincides with Eq.~(\ref{bfkl}). The conventional BFKL equation predicts 
that the small-$x$ rises of the gluon distribution function and of
structure functions involved in deep inelastic scattering are almost
$Q$-independent as indicated by Eq.~(\ref{bfk}). However, recent HERA data
of the structure function $F_2(x,Q^2)$ \cite{H1} exhibit a more sensitive
$Q$ dependence: the rise is rapider at larger $Q$. It is easy to observe
that the evolution kernel in Eq.~(\ref{mbf}) enhances the rise of
the gluon distribution function when $Q$ is large. It has been shown that
predictions from Eq.~(\ref{mbf}) are well consistent with the data 
\cite{L4}.

It is known that the BFKL rise of the gluon distribution function and
of the structure function $F_2$ violates the unitarity bound
$F_2 < {\rm const.}\times \ln^2(1/x)$. This is obvious from the solution of
$F$ to Eq.~(\ref{bfkl}), which grows like a power of $1/x$. Without
approximating $F(x/z,b,\mu)$ by $F(x,b,\mu)$, the term $1/(1-z)_+$ in 
$P_{gg}$ gives a destructive contribution to the evolution kernel because of
$F(x/z,b,\mu)\le F(x,b,\mu)$. It implies that the assumption of rapidity 
ordering overestimates real gluon contributions, which are responsible for 
the BFKL rise. Without assuming strong rapidity ordering, 
Eq.~(\ref{nunif}) becomes
\begin{eqnarray}
F(x,b,Q)&=&\Delta(x,b,Q,Q_0)F(x,b,Q_0)
\nonumber\\
& &+{\bar\alpha}_s\int_x^1 \frac{dz}{z}\int_{Q_0^2}^{Q^2}
\frac{d\mu^2}{\mu^2}\Delta(x,b,Q,\mu)F(x,b,\mu)
\nonumber\\
& &+{\bar\alpha}_s\int_x^1 \frac{dz}{z}\int_{Q_0^2}^{Q^2}
\frac{d\mu^2}{\mu^2}\Delta(x,b,Q,\mu)
\frac{F(x/z,b,\mu)-F(x,b,\mu)}{1-z},
\nonumber\\
& &
\label{nbku}
\end{eqnarray}
which contains a destructive term compared with Eq.~(\ref{nbk}).
Differentiating Eq.~(\ref{nbku}) with respect to $x$ and using 
Eq.~(\ref{sde}), we derive
\begin{eqnarray}
-x\frac{d}{d x}F(x,b) &=& -{\bar\alpha}_s\ln(Q_0^2 b^2)F(x,b)
\nonumber\\
& &+[{\bar\alpha}_s\ln(Q_0^2 b^2)]^2\int_x^1 \frac{dz}{z}
\frac{F(x/z,b)-F(x,b)}{1-z}\;,
\label{bfku}
\end{eqnarray}
where the variable $Q$ has been replaced by $1/b$ as in Eq.~(\ref{bfkl}). 
The correction term is of higher order, whose additional power
${\bar\alpha}_s\ln(Q_0^2b^2)$ comes from
the derivative of $\Delta$ with respect to $x$. 

Equation (\ref{bfku}) is similar to the modified BFKL equation with 
unitarity proposed in \cite{L5}. The behavior of $F(x/z,b)$ at $z\to x$ 
provides strong suppression to real gluon contributions, such that the 
power-law rise turns into a logarithmic rise, and unitarity is restored 
\cite{L5}. The reasoning is reviewed below. Inserting a guess solution 
$F\propto x^{-\lambda}$ into Eq.~(\ref{bfku}),
$\lambda$ being a parameter, we obtain
\begin{equation}
\lambda=S+S^2\int_x^1 \frac{dz}{z}\frac{z^\lambda-1}{1-z}\;,
\label{tb}
\end{equation}
with $S=-{\bar\alpha}_s\ln(Q_0^2 b^2)$. A solution of $\lambda$, 
$0< \lambda < S$, exists for a $x$ which is not too small, since the 
correction term, {\it i.e.}, the second term on the right-hand side of 
Eq.~(\ref{tb}), is negative and not large. That is, $F$ increases as a 
power of $x$, consistent with the results from the conventional  
BFKL equation. While the correction term diverges as $x\to 0$,
and no solution of $\lambda$ is allowed, implying that $F$ can not 
maintain the power-law rise at extremely small $x$. We then substitute
another guess $F|_{x\to 0}\propto \ln(1/x)$ with a milder rise into 
Eq.~(\ref{bfku}). In this case the correction term, increasing as
$\ln(1/x)$ roughly, cancels the first term $SF\propto \ln(1/x)$. The 
derivative term $-xdF/dx\propto 1$ is negligible. Therefore, 
Eq.~(\ref{bfku}) holds approximately. At last, we assume 
${\tilde F}|_{x\to 0}\propto$ const. as a test. It is easy to find that the 
first term becomes dominant, and the correction term and the derivative 
term vanish, {\it i.e.}, no const.$\not=0$ exists. These simple
investigations indicate that $F$ should increase as $\ln(1/x)$ at most,
instead of as a power of $1/x$, when $x$ approaches zero. For a rigorous 
numerical solution to the modified BFKL equation, which exhibits a 
logarithmic rise at $x\to 0$ explicitly, refer to \cite{L5}. In conclusion, 
the new evolution equation, predicting a rapid power-like rise of the 
gluon distribution function $F$ for small $x$ and a milder logarithmic 
rise for $x\to 0$, satisfies the requirement of unitarity. 

We emphasize that predictions from the CCFM equation, based on 
strong angular ordering of radiative gluons, do not satisfy the
unitarity bound. A careful investigation shows that the BFKL rise embedded
in the CCFM equation is attributed to the same origin as in the new
evolution equation \cite{L4,KLM}:
\begin{eqnarray}
{\bar\alpha}_s\int_x^1 \frac{dz}{z}
\int\frac{d^2q}{\pi q^2}\theta(Q-zq)
\Delta(k_T,Q)F(x,|{\bf k}_T+(1-z){\bf q}|,q)\;,
\label{ccfm1}
\end{eqnarray}
where $\Delta_S(zq,Q)$ and $\Delta_{NS}(z,q,k_T)$ have combined into
$\Delta(k_T,Q)$. The correction to the power-law rise also arises from the
term $1/(1-z)_+$:
\begin{eqnarray}
{\bar\alpha}_s\int_x^1 \frac{dz}{z}
\int\frac{d^2q}{\pi q^2}\theta(Q-zq)
\Delta_S(zq,Q)\frac{F(x/z,|{\bf k}_T+(1-z){\bf q}|,q)}{(1-z)_+}\;.
\label{ccfm2}
\end{eqnarray}
However, the Sudakov exponential $\Delta_S(zq,Q)$
in Eq.~(\ref{ccfm2}) suppresses the region with $z\to x \to 0$, 
such that the correction to the assumption of
rapidity ordering in the CCFM equation is not important enough to
turn the power-law rise into a logarithmic rise. This correction
is not suppressed by $\Delta(x,b,Q,\mu)$ at small $x$ in the new equation.

Compared with the CCFM equation, the new evolution equation bears other  
similarities and differences. They both contain the Sudakov exponentials, 
but the forms are different. The upper bounds of the evolution variables 
are the same ($Q$), but the lower bounds are $zq$ in the CCFM equation, 
and $\mu$ in the new equation. The longitudinal variable $\mu$ controls the 
magnitude of the Sudakov form factor by changing the partition of 
collinear enhancements between $\Delta(x,b,Q,\mu)$ and $F(x/z,b,\mu)$. 
While $zq$ from angular ordering changes the partition of collinear 
enhancements between $\Delta_S(zq,Q)$ and $\Delta_{NS}(z,q,k_T)$. 
There is no separation of $\Delta_{NS}$ from $\Delta_S$ in the new equation. 
As explained in Sec.~2, this smearing factor $\Delta_{NS}$ is not necessary,
since the $z\to 0$ pole does not exist.

\vskip 1.0cm

\centerline{\large\bf 5. Conclusion}
\vskip 0.5cm

In this paper we have derived a new unified evolution equation for the
gluon distribution function using the CS resummation
technique, which is appropriate for both the large and small $x$.
The features of this equation are summarized below. The infrared
cancellation between real and virtual gluon emissions is explicit
in both the Sudakov form factor and the splitting function. The Sudakov
form factor can take into account next-to-leading logarithmic summation 
systematically, including the running of the coupling constant. Hard virtual 
gluon emissions introduces a desired $Q$ dependence into the evolution 
kernel. The splitting function does not contain the non-Sudakov form factor 
compared to that in the CCFM equation. Predictions for the gluon 
distribution function satisfy the unitarity bound. Hence, the new unified 
evolution equation can be regarded as  an improved version of the CCFM 
equation.

In the future studies we shall explore the properties of this new 
equation in details, and results will be published elsewhere.

\vskip 0.5cm
This work was supported by the National Science Council of R.O.C. under 
Grant No. NSC87-2112-M-006-013.

\newpage
\centerline{\large \bf Figure Captions}
\vskip 0.5cm

\noindent
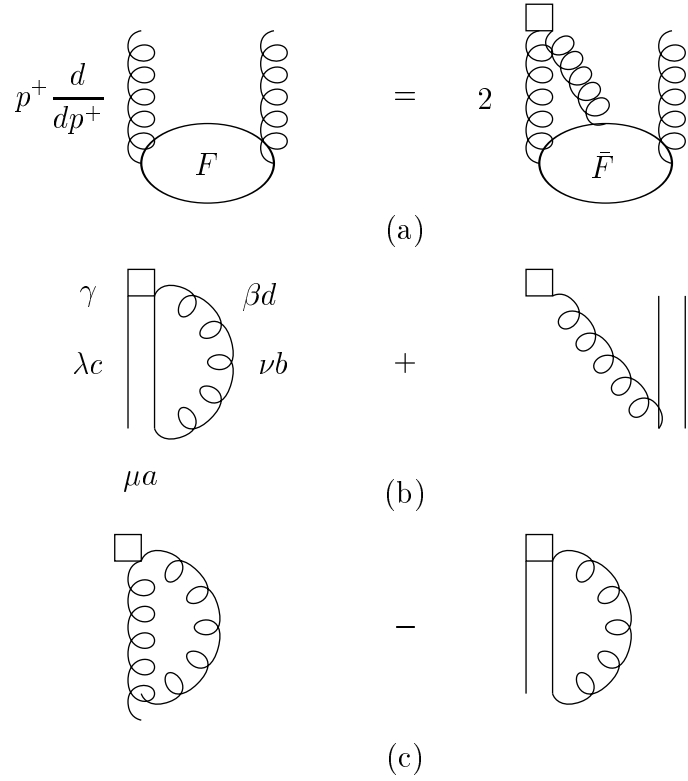
{\bf FIG. 1.} (a) The derivative $p^+dF/dp^+$ in the axial gauge. (b) The 
lowest-order subdiagrams for $K$. (c) The lowest-order subdiagrams for $G$.
\vskip 0.5cm

\noindent
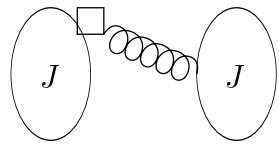
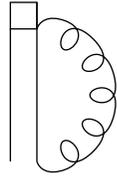
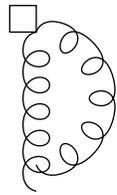
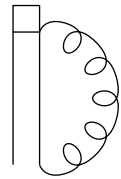
{\bf FIG. 2.} (a) The subdiagram containing the special vertex for the CCFM 
equation. (b) The lowest-order subdiagram for $K_J$. (c) The lowest-order 
subdiagrams for $G_J$.

\end{document}